# Schedulability Test for Soft Real-Time Systems under Multi-processor Environment by using an Earliest Deadline First Scheduling Algorithm

Jagbeer Singh
Dept. of Computer Science and Engineering
Gandhi Institute of Engg. & Tech.
Gunupur, Rayagada, India-765022
jsingh@giet.edu

## Abstract

This paper deals with the study of Earliest Deadline First (EDF) which is an optimal scheduling algorithm for uniprocessor real time systems use for scheduling the periodic task in soft real-time multiprocessor systems. In hard real-time systems, a significant disparity exists EDF-based schemes and RMA scheduling (which is the only known way of optimally scheduling recurrent real-time tasks on multiprocessors): on $M$ processors, all known EDF variants have utilization-based schedulability bounds of approximately $M/2$, while RMA algorithms can fully utilize all processors. This is unfortunate because EDF-based algorithms entail lower scheduling and task-migration overheads. In work on hard real-time systems, it has been shown that this disparity in Schedulability can be lessened by placing caps on per-task utilizations. Our main contribution is a new EDF-based scheme that ensures bounded deadline tardiness. In this scheme, per-task utilizations must be focused, but overall utilization need not be restricted. Our scheme should enable a wide range of soft real-time applications to be scheduled with *no constraints on total utilization*. Also propose techniques and heuristics that can be used to reduce tardiness as well as increase the efficiency of task.

## 1 Introduction

A real-time system is per definition correct if it performs the correct *function* at the correct *time*. Using real-time scheduling theory we can provide guarantees that each task in the system will meet its timing requirements [11][6][12], given that the basic assumptions, e.g., concerning task execution times and periodicity, are not violated at run-time. Real-time systems are those in which its correct operation not only depends on the logical results, but also on the time at which these results are produced. These are high complexity systems that are executed in environments such as: military process control, robotics, avionics systems, distributed systems and multimedia.

Real-time systems use scheduling algorithms to decide an *order of execution* of the tasks and an *amount of time* assigned for each task in the system so that no task (for hard real-time systems) or a minimum number of tasks (for soft real-time systems) misses their deadlines. In order to verify the fulfillment of the temporal constraints, real-time systems use different exact or inexact *schedulability tests*. The schedulability test decides if a given task set can be scheduled such that no tasks in the set miss their deadlines. *Exact schedulability tests* usually have high time complexities and may not be adequate for online admission control where the system has a large number of tasks or a dynamic workload. In contrast, inexact schedulability tests provide low complexity sufficient schedulability tests.

The first schedulability test known was introduced by Liu and Layland with the Rate Monotonic Scheduling Algorithm [Liu, 1973] (RM). Liu and Layland introduced the concept of achievable utilization factor to provide a low complexity test for deciding the schedulability of independent periodic and preemptable task sets executing on one processor. In Earliest Deadline First scheduling, at every scheduling point the task having the shortest deadline is taken up for scheduling. The basic principle of this algorithm is very intuitive and simple to understand. The schedulability test for EDF is also simple. A task is schedule under EDF, if and only if it satisfies the condition that total processor utilization ($U_i$) due to the task set is less than 1.

The most important attribute of real-time systems is that the correctness of such systems depends on not only the running results but also on the time at which results are produced. Real-time systems have to guarantee that all the strict timing requirements must be satisfied. In other words, real-time systems have timing requirements that must be guaranteed. Scheduling and schedulability analysis enables these guarantees to be provided.

Schedulability tests for general EDF systems with arbitrary relative deadlines can be sufficient or exact (necessary and sufficient). Sufficient tests are usually efficient but they are not powerful, many schedulable task sets are not judged to be schedulable. The simplest sufficient tests are utilization based and they have polynomial complexity, however we observed that nearly all the task sets which are randomly generated in our experiments [10] cannot be correctly evaluated by such tests. Exact tests can be performed by processor demand analysis, which calculates the processor demand of a task set at every absolute deadline to check if there is an overflow in a specified time interval. In such an interval, there could be a very large number of deadlines that need to be verified.

Real-time multiprocessor systems in contrast of designs range from single-chip architectures, with a modest number of processors, to large-scale signal-processing systems, such as synthetic-aperture radar systems. In recent years, scheduling techniques for such systems have received considerable attention. In an effort to catalogue these various techniques, Carpenter *et al.* [4] suggested the categorization shown in Table 1, which pertains to scheduling schemes for *periodic* or *sporadic* tasks systems. In such systems, each task is invoked repeatedly, and each such invocation is called a *job*. The table classifies scheduling schemes along two dimensions:

1. *Complexity of the priority mechanism. Along this dimension, scheduling disciplines are categorized according to whether task priorities are (i) static, (ii) dynamic but fixed within a job, or (iii) fully-dynamic. Common examples of each type include (i) rate-monotonic algorithm (RMA) [6], (ii) earliest-deadline-first(EDF) [6], and (iii) least-laxity-first(LLF) [8] scheduling.*

2. *Degree of migration allowed. Along this dimension, disciplines are ranked as follows: (i)*

*no migration (i.e., task partitioning), (ii) migration allowed, but only at job boundaries (i.e., dynamic partitioning at the job level), and (iii) unrestricted migration (i.e., jobs are also allowed to migrate).*

According to Table 1, scheduling algorithms from only one category can schedule tasks correctly with no utilization loss, namely, algorithms that allow full migration and use fully-dynamic priorities (the top right entry). The fact that it is possible for algorithms in this category to incur no utilization loss follows from work on scheduling algorithms that ensure proportionate fairness (RMAness) [3]. RMA algorithms break tasks into smaller uniform pieces called subtasks," which are then scheduled. The subtasks of a task may execute on any processor, i.e., tasks may migrate within jobs. Hence, RMA scheduling algorithms may suffer higher scheduling and migration overheads than other schemes. Thus, the other categories in Table 1 are still of interest. In four of the other categories, the term $\alpha$ represents a cap on individual task utilizations. Note that, if such a cap is not exploited, then the upper bound on schedulable utilization for *each* of the other categories is approximately $M/2$ lower. This is no accident: as shown in [4], *no* algorithm in these categories can successfully schedule all task systems with total utilization at most $B$ on $M$ processors, where $(M+1)/2 < B \le M$. Given the scheduling and migration overheads of RMA algorithms, the disparity in schedulability between RMA algorithms and those in other categories is somewhat disappointing.

| | 1: static | 2: job-level dynamic | 3: fully dynamic |
|---|---|---|---|
| 3: full migration | $\frac{M^2}{2M-1} \le U \le \frac{M+1}{2}$ | $U \ge M - \alpha(M-1)$, if $\alpha \le \frac{1}{2}$ ; $\frac{M^2}{2M-1} \le U \le \frac{M+1}{2}$, otherwise | $U = M$ |
| 2: restricted migration | $U \le \frac{M+1}{2}$ | $U \ge M - \alpha(M-1)$, if $\alpha \le \frac{1}{2}$ ; $M - \alpha(M-1) \le U \le \frac{M+1}{2}$, otherwise | $U \ge M - \alpha(M-1)$, if $\alpha \le \frac{1}{2}$ ; $M - \alpha(M-1) \le U \le \frac{M+1}{2}$, otherwise |
| 1: partitioned | $(\sqrt{2}-1)M \le U \le \frac{M+1}{1+2^{\frac{1}{M+1}}}$ | $U = \frac{\beta M+1}{\beta+1}$, where $\beta = \lfloor \frac{U}{\alpha} \rfloor$ | $U = \frac{M+1}{2}$ |

Table 1: Known lower and upper bounds on schedulable utilization (denoted $U$) for the different classes of preemptive scheduling algorithms.

Fortunately, as the table suggests, if individual task utilizations can be focused, then it is sometimes possible to significantly relax restrictions on total utilization. For example, in the entries in the middle column, as approaches 0, $U$ approaches $M$. This follows from work on multiprocessor EDF scheduling [1],[2],[7], which shows that an interesting "middle ground" exists between unrestricted EDF-based algorithms (which have upper bounds of approximately $M/2$ on schedulable utilization) and RMA algorithms (which have a schedulable utilization bound of $M$). In essence, establishing this middle ground involved addressing the following question: *if per-task utilizations are restricted, and if no deadlines can be missed, then what is the largest overall utilization that can be allowed?* In this paper, we approach this middle ground in a different way by addressing this question: *if per-task utilizations are restricted, but overall utilization is not, then by how much can deadlines be missed?* Our interest in this question stems from the increasing prevalence of applications such as networking, multimedia, and immersive graphics systems (to name a few) that have only soft real-time requirements.

The maximum tardiness that any task may experience in our scheme is dependent on the per-task utilization cap assumed| the lower the cap, the lower the tardiness threshold. Even with a cap as high as 0.5 (*half* of the capacity of one processor), reasonable tardiness bounds can be guaranteed for a significant percentage of task systems. (In contrast, if $\alpha = 0.5$ in the middle entry of Table 1, then approximately 50% of the system's overall capacity may be lost.) Hence, our scheme should enable a wide range of soft real-time applications to be scheduled in practice with *no constraints on total utilization*. In addition, when a job misses its deadline, we do *not* require a commensurate delay in the release of the next job of the same task. As a result, each task's required processor share is maintained in the long term.

The motivation for providing faster exact schedulability analysis for general EDF systems is two-fold. As part of the design process many different parameter profiles may need to be checked. An automated search may even be undertaken as part of the architectural definition of the system. An efficient but accurate schedulability scheme is therefore needed. The second requirement comes from online systems. During the run-time of a system there could be new tasks arrive that need (if possible) to be added to the task set. The system must recalculate schedulability online to decide whether to allow the new tasks to enter into the system. Such online admission control gives a much higher requirement for the performance of the schedulability test as the decisions have to be made in a very short time and should not occupy too much system resource. Our scheme has the additional advantage of limiting migration costs, even in comparison to other EDF-based schemes: only up to $M-1$ tasks, where $M$ is the number of processors, *ever* migrate, and those that do, do so only between jobs. As noted in [4], migrations between jobs should not be a serious concern in systems where little per-task state is carried over from one job to the next.

The rest of this paper is organized as follows. In Sec. 2, our system model is presented. In Sec. 3 our proposed algorithm is described and a tardiness bound is derived for it. Techniques and heuristics that can be used to reduce tardiness observed in practice are presented in Sec. 4. In Sec. 5, a simulation-based evaluation of our basic algorithm and proposed heuristics is presented. Finally, we conclude in Sec. 6.

## 2 System Model

A hard real-time system comprises a set of $n$ real-time tasks { $\tau_1$, $\tau_2$, $\tau_3$.., } each task consists of an infinite or finite stream of jobs or requests which must be completed before their deadlines. Let $\tau$ indicate any given task of the system. Each task can be periodic or sporadic.

*Periodic tasks.* All jobs of a periodic task $\tau$ have a regular interarrival time $T_i$, we call $T_i$ the period of the periodic task $\tau$. If a job for a periodic task $\tau$ arrives at time $t$ ,then the next job of task $\tau$ must arrive at $t + T_i$.

*Sporadic tasks*. The jobs of a sporadic task $\tau$ arrive irregularly, but they have a minimum interarrival time $T_i$ , we call $T_i$ the period of the sporadic task $\tau$ . If a job of a sporadic task $\tau_i$ arrives at $t$ , then the next job of task $\tau$ can arrive at any time at or after $t_i + T$ .

If there are periodic tasks in the system, since in realistic situations it is difficult to forecast or to handle the exact starting time of all tasks when a system starts up, the first job of each periodic task is assumed or arrives at the same time. Each job of task $\tau$ requires up to the same worst-case execution time which equals the task $\tau$ 's worst-case execution time $C_i$, and each job of task $\tau_i$ has the same relative deadline which equals the task $\tau$ 's relative deadline $D_i$. If a job of task $\tau_i$ arrives at time $t$, the required worst-case execution time $_i C$ must be completed in $D_i$ time units, and the absolute deadline of this job is $t + D_i$ . Each task could have release jitter, when a job of task $\tau$ arrives at time $t$ with the absolute deadline $t_i + D$ , it will be released for execution at the latest time $t_i + J$ (the actual release time can be early than $t_i + J$)[13].

At any time, an arrived job with a higher priority can preempt a lower priority job's execution. When a job completes its execution, the system chooses the pending job with the highest priority to execute. According to the EDF algorithm, the released job with the earliest absolute deadline is assigned the highest priority.

The following notation is used throughout the paper.

$C_i$ —the worst-case execution time of task $\tau_i$

$D_i$ —the relative deadline of task $\tau_i$

$T_i$ —the period of task $\tau_i$

$n$ —the number of tasks in the system or the task set

$U_i$ —the utilization of task $\tau_i$, and $U_i = C_i / T_i$

$U$ —the total utilization of the task set, and

$$U = \sum_{i=1}^{n} \frac{c_i}{T_i}$$

$J_i$ —the maximum release jitter of task $\tau_i$

$d_i$ —an absolute deadline of a job of task $\tau_i$

$r_i$ —a job (or a request) of task $\tau_i$

We consider the scheduling of a recurrent (periodic or sporadic) task system $\tau$ comprised of **N** tasks on **M** *identical* processors. The $k^{th}$ processor is denoted $P_k$, where $1 \leq k \leq M$. Each task $T_i$, where $1 \leq i \leq n$, is characterized by a *period* $p_i$, an *execution cost* $e_i \leq p_i$, and a *relative deadline* $d_i$. Each task $T_i$ is invoked at regular intervals, and each invocation is referred to as a *job* of $T_i$. The $k^{th}$ job of $T_i$ is denoted $T_{i,k}$. The first job may be invoked or *released* at any time at or after time zero and the release times of any two consecutive jobs of $T_i$ should differ by at least $p_i$ time units. If every two consecutive jobs of $T_i$ differ by exactly $p_i$ time units, then $T_i$ is said to be a *periodic* task; otherwise, $T_i$ is a *sporadic* task. Every job of $T_i$ has a worst-case execution requirement of $e_i$ time units and an absolute deadline given by the sum of its release time and its relative deadline, $d_i$. In this paper, we assume that $d_i = p_i$ holds, for all $i$. We sometimes use the notation $T_i(e_i, p_i)$ to concisely denote the execution cost and period of task $T_i$.

The *utilization* of task $T_i$ is denoted $u_i$ and is given by $e_i / p_i$. If $u_i \leq 1/2$, then $T_i$ is called a *light task.* In this paper, we assume that every task to be scheduled is light. Because a light task can consume up to half the capacity of a single processor, we do not expect this to be a restrictive assumption in practice. The *total utilization* of a task system $\tau$ is defined as $U_{sum}(\tau) = \sum_{i=1}^{n} u_i$. A task system is said to *fully utilize* the available processing capacity if its total utilization equals the number of processors ($M$). The maximum utilization of any task in $\tau$ is denoted $u_{\max}(\tau)$. A task system is *preemptive* if the execution of its jobs may be interrupted and resumed later. In this paper, we consider only preemptive scheduling policies. We also place no constraints on total utilization[13]

The jobs of a *soft* real-time task may occasionally miss their deadlines, if the amount by which a job misses its deadline, referred to as its *tardiness*, is bounded. Formally, the tardiness of a job $T_{i,j}$ in schedule $S$ is defined as *tardiness*$(T_{i,j}, S) = \max(0, t - t_a)$, where $t$ is the time at which $T_{i,j}$ completes executing in $S$ and $t_a$ is its absolute deadline. The tardiness of a task system $\tau$ under scheduling algorithm $A$, denoted *tardiness*$(\tau, A)$, is defined as the maximum tardiness of any job in $\tau$ under any schedule under $A$. If $\cdot$ is the maximum tardiness of any task system under $A$, then $A$ is said to *ensure a tardiness bound of* ·. Though tasks in a soft real-time system are allowed to have nonzero tardiness, we assume that *missed deadlines do not delay future job releases*. That is, if a job of a task misses its deadline, then the release time of the next job of that task remains unaltered. Of course, we assume that consecutive jobs of the same task cannot be scheduled in parallel. Thus, a missed deadline effectively reduces the interval over which the next job should be scheduled in order to meet its deadline.

Our goal in this paper is to derive an EDF-based multiprocessor scheduling scheme that ensures bounded tardiness. In a "pure" EDF scheme, jobs with earlier deadlines would (always) be given higher priority. In our scheme, this is usually the case, but (as explained later) certain tasks are treated specially and are prioritized using other rules. Because we do not delay future job releases when a deadline is missed, our scheme ensures (over the long term) that each task receives a processor share approximately equal to its utilization. Thus, it should be useful in settings where maintaining correct share allocations is more important than meeting every deadline. In addition, schemes that ensure bounded tardiness are useful in systems in which a utility function is defined for each task [5][15]is function specifies the value" or usefulness of the current job of a task as a function of time; beyond a job's deadline, its usefulness typically decays from a positive value to 0 or below. The amount of time after its deadline beyond which the completion of a job has no value implicitly specifies a tardiness threshold for the corresponding task.

# 3 Algorithms Feasible EDF

Let $m$ denote the number of processing nodes and $n$, ($n \geq m$) denote the number of Available tasks in a uniform parallel real-time system. Let $s_1, s_2, \dots s_m$ denote the computing capacity of available processing nodes indexed in a non-increasing manner: $s_j \geq s_j +1$ for all $j$, $1 < j < m$. We assume that all speeds are positive i.e. $s_j > 0$ for all j. In this section we are presenting five steps of *EFDF* algorithm. Obviously, each task which is picked for up execution is not considered for execution by other processors. Here we are giving following methods for our new approach:

1.  Perform a feasibility check to specify the task which has a chance to meet their deadline and put them into a set **A**, Put the remaining tasks into set **B**. We can partition the task set by any existing approach.

2.  Sort both task sets **A** and **B** according to their deadline in a non-descending order by using any of existing sorting algorithms. Let **k** denote the number of tasks in set **A**, i.e. the number of tasks that have the opportunity to meet their deadline.

3.  For all processor **j**, ($j \leq min(k,m)$) check whether a task which was last running on the **j**$^{th}$ processor is among the first $min(k,m)$ tasks of set **A**. If so assign it to the **j**$^{th}$ processor. At this point there might be some processors to which no task has been assigned yet.

4.  For all **j**, ($j \leq min(k,m)$) if no task is assigned to the **j**$^{th}$ processor , select the task with earliest deadline from remaining tasks of set A and assign it to the **j**$^{th}$ processor. If **k** $\geq$ **m**, each processor have a task to process and the algorithm is finished.

5.  If **k** < **m**, for all **j**, ($k < j \leq m$) assign the task with smallest deadline from **B** to the **j**$^{th}$ processor. The last step is optional and all the tasks from **B** will miss their deadlines.

In this section, we propose Algorithm Feasible EDF an EDF-based multiprocessor scheduling algorithm that ensures bounded tardiness for task systems whose per-task utilizations are at most 1/2. Feasible EDF does not place any restrictions on the total system utilization. Further, at most $M$ -1 task needs to be able to migrate, and each such task migrates between two processors, across job boundaries only. This has the benefit of lowering the number of tasks whose states need to be stored on a processor and the number of processors on which each task's state needs to be stored. Also, the runtime context of a job, which can be expected to be larger than that of a task, need not be transferred between processors.

Feasible EDF consists of two phases: an *distribution phase* and an *execution phase*. The distribution phase executes o²ine and consists of sequentially assigning each task to one or two processors. In the execution phase, jobs are scheduled for execution at runtime such that over reasonable intervals (as explained later), each task executes at a rate that is commensurate with its utilization. The two phases are explained in detail below. The following notation shall be used.

$S_{i,j}$ $\stackrel{def}{=}$ Percentage of $P_j$'s processing capacity

(expressed as a fraction) allocated to $T_i, 1 \leq i \leq n;\ 1 \leq j \leq M$. ($T_i$ is said to have a *share* of $S_{i,j}$ on $P_j$.) $\qquad(1)$

$F_{i,j}$ $\stackrel{def}{=}$ $\frac{s_{i,j}}{u_i}$, the fraction of $T_i$'s total execution

requirement that $P_j$ can handle, $1 \leq i \leq n;\ 1 \leq j \leq M$ $\qquad(2)$

## 3.1 Distribution Phase

Tasks assigned to two processors are called *migrating* tasks, while those assigned to only one processor is called *fixed* or *non-migrating* tasks. The Distribution phase represents a mapping of tasks to processors. Each task is assigned to either one or two processors. A fixed task $T_i$ is assigned a *share*, $s_{i,j}$, equal to its utilization $u_i$ on the only processor $P_j$ to which it is assigned. A migrating task has shares on both processors to which it is assigned. The sum of its shares equals its utilization. The distribution phase of Feasible EDF also ensures that at most two migrating tasks are assigned to any processor. In Fig. 1, a task-distribution algorithm, denoted Assign-Tasks, is given that satisfies the following properties for any task set $\tau$ with $u_{max}(\tau) \leq 1/2$ and $U_{sum}(\tau) \leq M$.

(P1) Each task is assigned shares on at most two processors only. A task's total share equals its utilization.

(P2) Each processor is assigned at most two migrating tasks only and may be assigned any number of fixed tasks.

(P3) The sum of the shares allocated to the tasks on any processor is at most one.

In the pseudo-code for this algorithm, the $i_{th}$ element $u[i]$ of the global array $u$ represents the utilization $u_i$ of task $T_i$, $s[i][j]$ denotes $s_{i,j}$ (as defined in (1)), array $p[i]$ contains the processor(s) to which task $i$ is assigned, and arrays $m[i]$ and $f[i]$ denote the migrating tasks and fixed tasks assigned to processor $i$, respectively. Note that $p[i]$ and $m[i]$ are each vectors of size two.

## 3.2 Previous Results on Exact Schedulability Analysis and execution phase

We first argue that bounding total demand may not be possible if the jobs of migrating tasks are allowed to miss their deadlines. In order to analyze a scheduling algorithm and for the algorithm to guarantee bounded tardiness, it should be possible to bound the total *demand* for execution time by all tasks on each processor over well defined time intervals.

For a fixed task, we merely need to decide when to schedule each of its jobs on its (only) assigned processor. For a migrating task, we must decide both *when* and *where* its jobs should execute Having devised a way of assigning tasks to processors, the next step is to devise an online scheduling algorithm that is easy to analyze and ensures bounded tardiness.. Before describing our scheduling algorithm, we discuss some considerations that led to its design.

In the worst case, the second processor may be forced to idle. The tardiness of the second job may also impact the timeliness of fixed tasks and other migrating tasks assigned to the same processor, which in turn may lead to dead-line misses of both fixed and migrating tasks on other processors or unnecessary idling on other processors. Because a deadline miss of a job does not lead to a postponement of the release times of subsequent jobs of the same task, and because two jobs of a task may not execute in parallel, the tardiness of a job of a migrating task executing on one processor can affect the tardiness of its successor job, which may otherwise execute in a timely manner on a second processor.

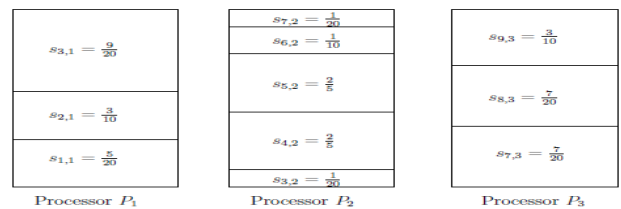

Figure 2: Example task distribution on three processors using Algorithm Assign-Tasks.

As a result, a set of dependencies is created among the jobs of migrating tasks, resulting in an intricate linkage among processors that complicates scheduling analysis. It is unclear how per-processor demand can be precisely bounded when activities on different processors become interlinked.

Let we consider for a concrete example that reveals this linkage among processors. Consider task set $\tau$ introduced earlier, with task distributions and processor shares shown in Fig. 2. For simplicity, assume that the execution of the jobs of a migrating task alternate between the two processors to which the task is assigned. $T_3$ releases its first job on $P_1$, while $T_7$ releases its first job on $P_3$. (We are assuming such a naive distribution pattern to illustrate the processor linkage using a short segment of a real schedule. Such a linkage occurs even with an intelligent job-distribution pattern if migrating tasks miss their deadlines.) A complete schedule up to time 27, with the jobs assigned to each processor scheduled using EDF, is shown in Fig. 3.

In Fig. 3, the sixth job of the migrating task $T_3$ misses its deadline (at time 12) on $P_2$ and completes executing at time 14. This prevents the next job of $T_3$ released on $P_1$ from being scheduled until time 14 and it misses its deadline. Recall that a deadline miss does not cause future job releases to be postponed, thus the seventh job of $T_3$ is released at time 12 and has a deadline at time 14.

**Per-processor scheduling rules** The jobs assigned to a processor are scheduled independently of other processors, and on each processor, migrating tasks are statically prioritized Feasible EDF eliminates this linkage among processors by ensuring that migrating tasks do not miss their deadlines. Jobs of migrating tasks are assigned to processors using static rules that are independent of runtime dynamics.

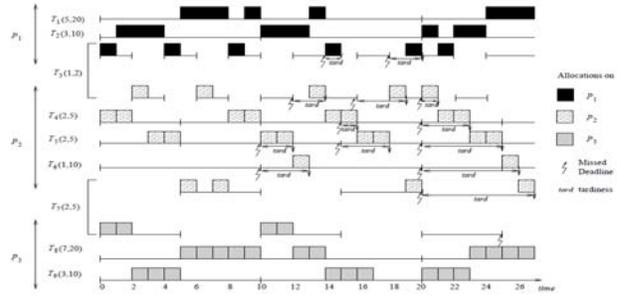

Figure 3: Illustration of processor linkage.

over fixed tasks. Jobs within each task class are scheduled using EDF, which is optimal on uniprocessors. This priority scheme, together with the restriction that migrating tasks have utilizations at most 1/2 and the task distribution property (from P2)) that there are at most two migrating tasks per processor, ensures that migrating tasks never miss their deadlines. Therefore, the jobs of migrating tasks executing on different processors do not impact one another, and each processor can be analyzed independently. Thus, the multiprocessor scheduling analysis problem at hand is transformed into a simpler uniprocessor one.

Feasible EDF follows a job distribution pattern that prevents over utilization in the long run by ensuring that over well-defined time intervals (explained later), the demand due to a migrating task on each processor is in accordance with its allocated share on that processor. In the description of Feasible EDF, we are left with defining rules that map jobs of migrating tasks to processors. A naive distribution of the jobs of a migrating task to its processors can cause an over utilization on one of its assigned processors. For example, consider the migrating task $T_7(2, 5)$ in the example above. $T_7$ has a share of $s_{7,2} = 1/20$ on $P_2$ and $s_{7,3} = 7/20$ on $P_3$. Also, $f_{7,2} = s_{7,2} / u_7 = 1/8$ and $f_{7,3} = s_{7,3} / u_7 = 7/8$, which imply that $P_2$ and $P_3$ should be capable of executing 1/8 and 7/8 of the workload of $T_7$, respectively. Our goal is to devise a job distribution pattern that would ensure that in the long run, the fraction of a migrating task $T_i$'s workload executed on $P_j$ is close to $f_{i,j}$. One such job distribution pattern for $T_7$ over interval

## .3.2.1 Excursion of RMA Scheduling

It is a mathematical model that contains a calculated simulation of periods in a closed system, where round-robin and time-sharing schedulers fail to meet the scheduling needs otherwise. Rate monotonic scheduling looks at a run modeling of all threads in the system and determines how much time is needed to meet the guarantees for the set of threads in question.

Liu & Layland (1973) proved that for a set of $n$ periodic tasks with unique periods, a feasible schedule that will always meet deadlines exists if the CPU utilization is below a specific bound (depending on the number of tasks). The schedulability test for **RMS** is:

$$U = \sum_{i=1}^{n} \frac{C_i}{T_i} \leq n(2^{1/n} - 1)$$

Where $C_i$ is the computation time, $T_i$ is the release period (with deadline one period later), and $n$ is the number of processes to be scheduled. For example $U \leq 0.8284$ for $n = 2$. When the number of processes tends towards infinity this expression will tend towards:

$$\lim_{n \to \infty} n(\sqrt[n]{2} - 1) = \ln 2 \approx 0.693147.$$

So a rough estimate is that **RMS** in the general case can meet all the deadlines if CPU utilization is 69.3%. The other 30.7% of the CPU can be dedicated to lower-priority non real-time tasks. It is known that a randomly generated periodic task system will meet all deadlines when the utilization is 85% or less,[3] however this fact depends on knowing the exact task statistics (periods, deadlines) which cannot be guaranteed for all task sets. The *rate monotonic* priority assignment is *optimal* meaning that if any *static priority* scheduling algorithm can meet all the deadlines, then the *rate monotonic* algorithm can too. The deadline-monotonic scheduling algorithm is also optimal with equal periods and deadlines, in fact in this case the algorithms are identical; in addition, deadline monotonic scheduling is optimal when deadlines are less than periods. Currently, RMA scheduling [3] is the only known way of *optimally* scheduling recurrent real-time task systems on multiprocessors. In RMA scheduling terminology, each task $T$ has an integer execution cost $T.e$ and an integer period $T.p \geq T.e$. The utilization of $T$, $T.e / T.p$, is also referred to as the *weight* of $T$ and is denoted $wt(T)$. (Note that in the context of RMA scheduling, tasks are denoted using upper-case letters without subscripts.)

RMA algorithms allocate processor time in discrete quanta that are uniform in size. Assuming that a quantum is one time unit in duration, the interval $[t, t+1)$, where $t$ is a non-negative integer, is referred to as *slot t*. At most one task may execute on each processor in each slot, and each task may execute on at most one processor only in every slot. The sequence of allocation decisions over time slots defines a *schedule S*. Formally, $S : \tau \times \mathbb{N} \mapsto \{0, 1\}$. $S(T, t) = 1$ iff $T$ is scheduled in slot $t$.

The notion of a RMA schedule for a periodic task $T$ is defined by comparing such a schedule to an ideal fluid schedule, which allocates $wt(T)$ processor time to $T$ in each slot. Deviation from the allocation in a fluid schedule is formally captured by the concept of *lag*. Formally, the *lag of task T at time t* in schedule $S$ is the difference between the total allocations to $T$ in a fluid schedule and $S$ in the interval $[0, t)$, *i.e.*,

$$lag(T,t,S) = wt(T) . t - \sum_{u=0}^{t-1} S(T,u) \quad (3)$$

A schedule $S$ is said to be *RMA* iff

$$(\forall T, t :: -1 < lag(T, t, S) < 1) \quad (4)$$

holds. Informally, the allocation error associated with each task must always be less than one quantum. The above constraints on lag have the effect of breaking task $T$ into a potentially infinite sequence of quantum length *subtasks*. The $i_{th}$ subtask of $T$ is denoted $T_i$, where $i \geq 1$. (In the context of RMA scheduling, $T_i$ does not denote the $i_{th}$ task, but the $i_{th}$ subtask of task $T$.) Each subtask $T_i$ is associated with a *pseudo-release* $r(T_i)$ and a *pseudo-deadline* $d(T_i)$ defined as follows:

$$r(T_i) = \left\lfloor \frac{i-1}{wt(T)} \right\rfloor \quad (5)$$

$$d(T_i) = \left\lfloor \frac{i}{wt(T)} \right\rfloor \quad (6)$$

To satisfy (4), $T_i$ must be scheduled in the interval $w(T_i) = [r(T_i), d(T_i))$, termed its *window*. Fig. 5(a) shows the windows of the first job of a periodic task with weight 3/7. In this example, $r(T_1) = 0$, $d(T_1) = 3$, and $w(T_1) = [0, 3)$ hold for subtask $T_1$.

We next define the notion of a *complementary task*, which is used to guide the sequence in which the jobs of a migratory task are assigned to its processors.

**Definition 1:** With the below introduction to RMA scheduling, we are now ready to present the details of distributing the jobs of a migrating task between its processors.

Tasks $T$ and $U$ are *complementary* iff $wt(U) = 1 - wt(T)$. Tasks T and U shown in Fig. are complementary to one another. A partial RMA schedule for these two tasks on one processor, in which the subtasks of T are always scheduled in the last slot of their windows and those of U in the first slot, is also shown. We call such a schedule a complementary schedule. It is easy to show that such a schedule is always possible for two complementary periodic tasks.

### 3.2.2 Rules for Jobs of Migrating Tasks

Let $T_i$ be any migrating periodic task (we later relax the assumption that $T_i$ is periodic) that is assigned shares $s_{i,j}$ and $s_{i,j+1}$ on processors $P_j$ and $P_{j+1}$, respectively. (Note that every migrating task is assigned shares on two consecutive processors by Assign-Tasks.) As explained earlier, $f_{i,j}$ and $f_{i,j+1}$ (given by (2)) denote the fraction of the workload (*i.e.*, the total execution requirement) of $T$ that should be executed on $P_j$ and $P_{j+1}$, respectively, in the long run. By (P1), the total share allocated to $T_i$ on $P_j$ and $P_{j+1}$ is $ui$. Hence, by (2), it follows that

$$f_{i,j} + f_{i,j+1} = 1 \qquad (7)$$

Assuming that the execution cost and period of every task are rational numbers (that can be expressed as a ratio of two integers), $u_i$, $s_{i,j}$, and hence, $f_{i,j}$ and $f_{i,j+1}$ are also rational numbers. Let $f_{i,j} = x_{i,j} / y_i$, where $x_{i,j}$ and $y_i$ are positive integers that are relatively prime. Then, by (7), it follows that $f_{i,j+1} = y_i - x_{i,j} / y_i$. Therefore, one way of distributing the workload of $T_i$ between $P_j$ and $P_{j+1}$ that is commensurate with the shares of $T_i$ on the two processors would be to assign $x_{i,j}$ out of every $y_i$ jobs to $P_j$ and the remaining jobs to $P_{j+1}$.

We borrow from the aforementioned concepts of RMA scheduling to guide in the distribution of jobs. If we let $f_{i,j}$ and $f_{i,j+1}$ denote the weights of two fictitious RMA tasks, $V$ and $W$, and let a quantum span $p_i$ time units, then the following analogy can be made between the jobs of the migrating task $T_i$ and the subtasks of the fictitious tasks $V$ and $W$. First, slot $s$ represents the interval in which the $(s + 1)^{st}$ job of $T_i$, which is released at the beginning of slot $s$, needs to be scheduled. (Recall that slots are numbered starting from 0.) Next, subtask $V_g$ represents the $g^{th}$ job of the jobs of $T_i$ assigned to $P_j$ ; similarly, subtask $W_h$ represents the $h^{th}$ job of the jobs of $T_i$ assigned to $P_{j+1}$.

RMA tasks $V$ and $W$ are complementary. Therefore, a complementary schedule for $V$ and $W$ in which the subtasks of $V$ are scheduled in the first slot of their windows and those of $W$ in the last slot of their windows is feasible. Accordingly, we consider a job distribution policy in which the job of $T_i$ corresponding to the first slot in the window of subtask $V_g$ is assigned as the $g^{th}$ job of $T_i$ to $P_j$ and the job of $T_i$ corresponding to the last slot in the window of subtask $W_h$ is assigned as the $h^{th}$ job of $T_i$ to $P_{j+1}$, for all $g$ and $h$. This policy satisfies the following property.

(A) Each job of $T_i$ is assigned to exactly one of $P_j$ and $P_{j+1}$.

Fig. 6(a) shows a complementary schedule for the RMA tasks that represent the rates at which the jobs of task $T_7$ in the example we have been considering should be assigned to $P_2$ and $P3$. Here, tasks $V$ and $W$ are of weights $f_{7,2} = 1/8$ and $f_{7,3} = 7/8$, respectively. A job distribution based on this schedule will assign the first of jobs $8k + 1$ through $8(k + 1)$ to $P_2$ and the remaining seven jobs to $P_3$, for all $k$.

Thus far in our discussion, in order to simplify the exposition, we assumed that the job releases of task $T_i$ are periodic. However, note that the job distribution given by (8) is independent of "real" time and is based on the job number only. Hence, assigning jobs using (8) should be sufficient to ensure (A) even when $T_i$ is sporadic Here, we assume that $T_7$ is a sporadic task, whose sixth job release is delayed by 11 time units to time 36 from time 25. As far as $T_7$ is concerned, the interval [25, 36) is "frozen" and the job distribution resumes at time 36. As indicated in the figure, in any such interval in which activity is suspended for a migrating task $T_i$, no jobs of $T_i$ are released. Furthermore, the deadlines of all jobs of $T_i$ released before the frozen interval fall at or before the beginning of the interval.

We next prove a property that bounds from above the number of jobs of a migrating task assigned to each of its processors by the job distribution rule given by previously.

**Lemma 1** *Let $T_i$ be a migrating task that is assigned to processors $P_j$ and $P_{j+1}$. The number of jobs out of any consecutive $l \geq 0$ jobs of $T_i$ that are assigned to $P_j$ and $P_{j+1}$ is at most $\lceil l.f_{i,j} \rceil$ and $\lceil l.f_{i,j+1} \rceil$, respectively.*

The tardiness bound given by Theorem 1 is directly proportional to the execution costs of the migrating tasks and the shares assigned to them. This bound could be high if the share of each migrating task is close to 1/2. However, because all tasks are light, in practice the sum of the shares of the migrating tasks assigned to a processor can be expected to be less than 1/2. Theorem 1 also suggests that the tardiness that results in practice could be reduced by choosing the set of migrating tasks carefully. Tardiness can also be reduced by distributing smaller pieces of works of migrating tasks than entire jobs. Some such techniques are discussed in the next section.

## 4 Delay Drop Techniques for FEDF

We consider the technique of *period transformation* [9] as a way of distributing the execution of jobs of migrating tasks more evenly over their periods in order to reduce the tardiness of fixed tasks. The problem of assigning tasks to processors such that the tardiness bound given by (21) is minimized is a combinatorial optimization problem with complexity that is exponential in the number of tasks. Hence, in this section, we propose methods and heuristics that can lower tardiness. We also propose task distribution heuristics that can reduce the fraction of a processor's capacity consumed by migrating tasks. The tardiness bound of Feasible EDF given by Theorem 1 is in multiples of the execution costs of migrating tasks. This is a direct consequence of statically prioritizing migrating tasks over fixed tasks and the overload (in terms of the number of jobs) that a migrating task may place on a processor over short intervals. The deleterious effect of this approach on jobs of fixed tasks can be mitigated by "slicing" each job of a migrating task into *sub-jobs* that have lower execution costs, assigning appropriate deadlines to the sub-jobs, and distributing and scheduling sub-jobs in the place of whole jobs. For example, every job of a task with an execution cost of 4 time units and relative deadline of 10 time units can be sliced into two sub-jobs with execution cost and relative deadline of 2 and 5, respectively, per sub-job, or four sub-jobs with an execution cost of 1 and relative deadline of 2.5, per sub-job. Such a job-slicing approach, termed *period transformation*, was proposed by Sha and Goodman [9] in the context of RM scheduling on uniprocessors. Their purpose was to boost the priority of tasks that have larger periods, but are more important than some other tasks with shorter periods, and thus ensure that the more important tasks do not miss deadlines under overloads. However, with the job-slicing approach under Feasible EDF, it may be necessary to migrate a job between its processors, and Feasible EDF loses the property that a task that migrates does so only across job boundaries. Thus, this approach presents a trade-off between tardiness and migration overhead.

**Task-distribution heuristics.** Another way of lowering the actual tardiness observed in practice would be to lower the total share $s_{mk,1,k} + s_{mk,2,k}$ assigned to the migrating tasks on any processor $P_k$. In the task distribution algorithm Assign-Tasks of Fig. 1, if a low utilization-task is ordered between two high utilization tasks, then it is possible that $s_{mk,1,k} + s_{mk,2,k}$ is arbitrarily close to one. For example, consider tasks $T_{i-1}$, $T_i$, and $T_{i+1}$ with utilizations $1-\varepsilon/2$, $2\varepsilon$, and $1-\varepsilon/2$, respectively, and a task distribution wherein $T_{i-1}$ and $T_{i+1}$ are the migrating tasks of $P_k$ with shares of $1-2\varepsilon/2$ each, and $T_i$ is the only fixed task on $P_k$. Such an distribution, which can delay $Ti$ excessively if the periods of $T_{i-1}$ and $T_{i+1}$ are large, can be easily avoided by ordering tasks by (monotonically) decreasing utilization prior to the distribution phase. Note that with tasks ordered by decreasing utilization, of all the tasks not yet assigned to processors, the one with the highest utilization is always chosen as the next migrating task. Hence, we call this distribution scheme *highest utilization first*, or HUF. An alternative *lowest utilization first*, or LUF, scheme can be defined that assigns fixed tasks in the order of (monotonically) decreasing utilization, but chooses the task with the lowest utilization of all the unassigned tasks as the next migrating task. Such an distribution can be accomplished using the following procedure when a migrating task needs to be chosen: traverse the unassigned task array in reverse order starting from the task with the lowest utilization and choose the first task whose utilization is at least the capacity available in the current processor. In general, this scheme can be expected to lower the shares of migrating tasks. However, because the unassigned tasks have to be scanned each time a migrating task is chosen, the time complexity of this scheme increases to $O(NM)$ (from $O(N)$). This complexity can be reduced to $O(M \log N)$ by adopting a binary search strategy.

**Counting non-light responsibilities** The primary reason for restricting all tasks to be light is to prevent the total utilization $u_i + u_j$ of the two migrating tasks $T_i$ and $T_j$ assigned to a processor from exceeding one. (As already noted, ensuring that migrating tasks do not miss their deadlines may not be possible otherwise.) However, if the number of non-light tasks is small in comparison to the number of light tasks, then it may be possible to assign tasks to processors without assigning two migrating tasks with total utilization exceeding one to the same processor. In the simulation experiments discussed in Sec. 5, with no restrictions on per-task utilizations, the LUF approach could successfully assign approximately 78% of the one million randomly-generated task sets on 4 processors. The success ratio dropped to approximately one-half when the number of processors increased to 16.

# 5 Experimental Evaluations

In this section, we describe the results of three sets of simulation experiments conducted using randomly generated task sets to evaluate Feasible EDF and the heuristics described in Sec. 4.

The experiments in the first set evaluate the various task distribution heuristics for varying numbers of processors, $M$, and varying maximum values of per-task utilization, $u_{\max}$. For each $M$ and $u_{\max}$, 1,000,000 task sets were generated. Each task set was generated as follows:

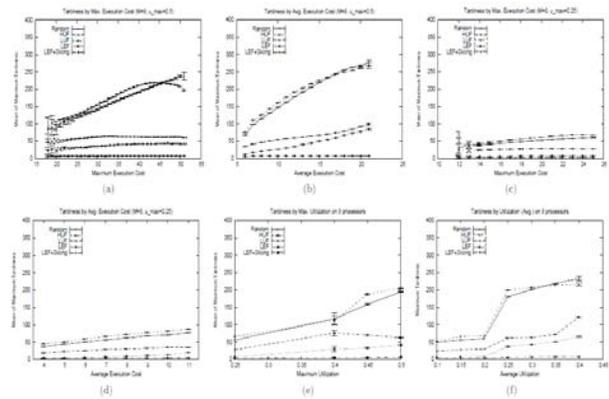

Figure 4: Comparison of different task distribution heuristics. Tardiness for $M = 8$ and $u_{\max} = 0.5$ by (a) $e_{\max}$ and (b) $e_{avg}$. Tardiness for $M = 8$ and $u_{\max} = 0.25$ by (c) $e_{\max}$ and (d) $e_{avg}$. Tardiness for $M = 8$ by (e) $u_{\max}$ and (f) $u_{avg}$.

New tasks were added to $\tau$ as long as the total utilization of $\tau$ was less than $M$. For each new task $Ti$, first, its period $pi$ was generated as a uniform random number in the range $[1; 100]$; then, its execution cost was chosen randomly in the range $[1{=}u_{max}; u_{\max} \notin pi]$. The last task was generated such that the total utilization of $\tau$ exactly equaled $M$. The generated task sets were classified in four different ways: (i) by the maximum execution cost of any task in a task set, $e_{\max}$, (ii) by the average execution cost of a task set, $e_{avg}$ (iii) by the maximum utilization of any task in a task set, $u_{\max}$, and (iv) by the average utilization of a task set, $u_{avg}$. The tardiness bound given by (21) was computed for each task set under a random task distribution and also under heuristics HUF, LUF, and LEF. The average value of the tardiness bound for task sets in each group under each classification and heuristic was then computed. The results for the groups classified by $e_{max}$ and $e_{avg}$ for $M = 8$ and $u_{max} = 0.5$ are shown in insets (a) and (b), respectively, of Fig. 4 Insets (c) and (d) contain the results under the same classifications for $M = 8$ and $u_{max} = 0.25$. (99% confidence intervals are also shown in these plots.) Results for classification by $u_{max}$ and $u_{avg}$ are given in insets (e) and (f), respectively. (It is somewhat difficult to distinguish the plots in the figures. Mostly, the orders in the legend and the plots coincide.)

The plots show that LEF guarantees the minimum tardiness of the four task-distribution approaches. Tardiness is quite low (approximately 8 time units mostly) under LEF for $u_{max} = 0.25$ (insets (c), (d), and (e)), which suggests that LEF may be a reasonable strategy for such task systems. Tardiness increases with increasing $u$max, but is still a reasonable value of 25 time units only for $e_{avg} \cdot 10$ when $u_{max} = 0.5$. However, for $e_{avg} = 20$, tardiness exceeds 75 time units, which may not be acceptable. For such systems, tardiness can be reduced by using the job-slicing approach, at the cost of increased migration overhead.

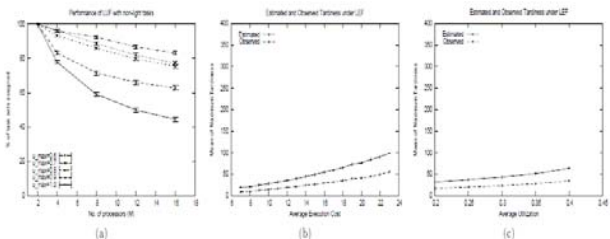

figure 5 (a) Percentage of randomly-generated task sets with non-light tasks successfully assigned by the LUF heuristic. (b) & (c) Comparison of estimated and observed tardiness under Feasible EDF-LEF by (b) average execution cost and (c) average utilization.

Therefore, in an attempt to determine the reduction possible with the job-slicing approach, we also computed the tardiness bound under LEF assuming that each job of a migrating task is sliced into sub-jobs with execution costs in the range $[1, 2)$. This bound is also plotted in the figures referred to above. For $M > 4$ and $u_{max} = 0.5$, we found the bound to settle to approximately $7\{8$ time units, regardless of the execution costs and individual task utilizations. (When $u_{max} = 0.25$, tardiness is $1\{2$ time units only under LEF with job slicing.) In our experiments, on average, a seven-fold decrease in tardiness was observed with job slicing with a granularity of one to two time units per sub-job. However, a commensurate increase in the number of migrations is also inevitable.

The second set of experiments evaluates the different heuristics in their ability to successfully assign task sets that contain non-light tasks also. Task sets were generated using the same procedure as that described for the first set of experiments above, except that $u_{max}$ was varied between $0.6$ and $1.0$ in steps of $0.1$. All of the four approaches could assign 100% of the task sets generated for $M = 2$, as expected. However, for higher values of $M$, the success ratio plummeted for all but the LUF approach. The percentage of task sets that LUF could successfully assign for varying $M$ and $u_{max}$ is shown in Fig. 5a). The third set of experiments was designed to evaluate the pessimism in the tardiness bound of (21). 300,000 task sets were generated with $u_{max} = 0.5$ and $U_{sum} = 8$. The tardiness bound estimated by (21) under the LEF task distribution heuristic was computed for each task set. A schedule under Feasible EDF-LEF for 100,000 time units was also generated for each task set and the actual maximum tardiness observed was noted. (The time limit of 100,000 was determined by trial-and-error as an upper bound on the time within which tardiness converged for the tasks sets generated.) Plots of the average of the estimated and observed values for tasks grouped by $e_{avg}$ and $u_{avg}$ are shown in insets (b) and (c) of Fig. 8, respectively. In general, we found that actual tardiness is only approximately half of the estimated value.

# 6 Conclusion and future scope

We have only taken a first step towards understanding tardiness under EDF-based algorithms on multiprocessors and have not addressed all practical issues concerned.

Foremost, the migration overhead of job slicing would translate into inflated execution costs for migrating tasks, and to an eventual loss of schedulable utilization. Hence, an iterative procedure for slicing jobs optimally may be essential. Next, our assumption that arbitrary task distributions are possible may not be true if tasks are not independent.wehave proposed a new algorithm, Feasible EDF, which is based on EDF, for scheduling recurrent soft real time task systems on multiprocessors, and have derived a tardiness bound that can be guaranteed under it. Our algorithm places no restrictions on the total system utilization, but requires per-task utilizations to be at most one-half of a processor's capacity. This restriction is very liberal, and hence, our algorithm can be expected to be sufficient for scheduling a large percentage of soft real-time applications. Though a global EDF algorithm, with no restrictions on migration, would appear to be capable of guaranteeing a lower tardiness bound than Feasible EDF, we have so far not been able to derive a non-trivial bound under it. In fact, we conjecture that a severe restriction on total utilization may be necessary, in addition to per-task utilization restrictions, to guarantee a non-trivial tardiness bound under unrestricted EDF. We believe that such studies should be conducted regularly by collecting data continuously so that skill demand patterns can be understood properly. This understanding can lead to informed curricula design that can prepare graduates equipped with necessary skills for employment. Once such studies are carried out, students can use the findings to select courses that focus on those skills which are in demand. Academic institutions can use the findings so that those skills in demand can be taken into account during curriculum design. As an advance to our work, in future, we have desire to work on different deployment approaches by developing more strong and innovative algorithms to solve the time complexity of Earliest Deadline First. Moreover, as our proposed algorithm is a generalized one, we have planned to expand our idea in the field of Real Time System existing Rate Monotonic Algorithm for calculating minimum Time Complexity. Moreover, we have aim to explore some more methodologies to implement the concept of this paper in real world and also explore for Fault Tolerance Task Scheduling Algorithms to finding the Task Dependency in single processor or multiprocessor system for reducing the time for fault also reduce the risk for fault and damage.